\documentclass[aps,amssymb,pra,amsmath,11pt,no title page,tightenlines]{revtex4}

\usepackage{epsfig}
\usepackage{bm}
\usepackage{ulem}

\def\br{{\bm r}}

\def\brt{\br,t}
\def\bp{{\bm p}}
\def\bv{{\bm v}}

\begin{document}

\title{Finite temperature vortex dynamics in Bose Einstein condensates}
\author{B.~Jackson$^{1}$\footnote{Brian Jackson died on 30 August 2007.},
N.~P.~Proukakis$^{1}$,
C.~F.~Barenghi$^{1}$ and E.~Zaremba$^{2}$}
\affiliation{$^1$ School of Mathematics and Statistics, University of
 Newcastle upon Tyne, NE1 7RU, United Kingdom,\\
$^2$Department of Physics, Engineering Physics and Astronomy,
Queen's University, Kingston, Ontario,
 Canada K7L 3N6,\\}

\begin{abstract}
We study the decay of vortices in Bose-Einstein condensates at finite
temperatures by means of the
Zaremba-Nikuni--Griffin formalism, in which the condensate is modelled by a Gross--Pitaevskii
equation, which is coupled to a Boltzmann kinetic equation for
the thermal cloud. 
At finite temperature, an off-centred vortex in a harmonically trapped 
pancake--shaped condensate decays by spiralling out towards the edge 
of the condensate. 
This decay, which depends heavily on temperature and atomic collisions, agrees with that predicted by the Hall--Vinen phenomenological model of friction force, which is used to 
describe quantised vorticity in superfluid systems. 
Our result thus clarifies the microscopic origin of the friction and provides an ab initio determination of its value. 
\end{abstract}
          
\date{\today}


\maketitle

\section{Introduction}
 
The dynamics of Bose-Einstein condensates at finite temperature presents an interesting
problem in the study of ultra cold Bose gases.
In most experiments, such systems are only partially condensed,
with the non-condensed thermal cloud providing 
a source of dissipation and leading to 
damping of structures, such as collective modes 
\cite{jin97,stamper-kurn97,marago01,chevy02}, 
solitons \cite{burger99} and vortices 
\cite{rosenbusch02,abo-shaeer02}.  
Several approaches have been developed to describe these systems,
including generalised mean field treatments \cite{Griffin_Popov,MF_Prouk,zaremba99,MF_JILA,MF_Milena,MF_Bijlsma}, 
number-conserving approaches \cite{NC_Gardiner,NC_Castin,NC_Morgan},
classical field theory 
\cite{Classical_Svistunov,Classical_PGPE,Classical_Review},
and stochastic approaches 
\cite{Stochastic_Gardiner,Stochastic_Stoof,Stochastic_Davis},
as recently reviewed by two of the authors \cite{Review_NickBrian},
who give a more complete list of references.
Although the 
underlying theory is well understood, the implementation of models
which can be actually solved in specific contexts has proven to be 
a considerable challenge, with the majority of treatments to date
assuming the thermal cloud is homogeneous and static.

In this paper we use the formalism of Zaremba, Nikuni and Griffin 
(ZNG) \cite{zaremba99}. The ZNG theory is a kinetic approach in which
a generalised Gross--Pitaevskii (GP) equation
for the condensate order parameter is coupled to a Boltzmann
equation for the thermal cloud. 
These equations have already been solved numerically
\cite{jackson02a}, and applied to the study
of collective modes \cite{jackson01,jackson02b,jackson02c},
the hydrodynamic regime \cite{hydro_1,hydro_2}
and the decay of dark solitons \cite{jackson07}, demonstrating 
good agreement with experiments.

\begin{figure}[b]
\centering \scalebox{0.2}
 {\includegraphics{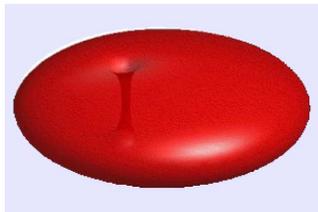}}
 \caption{(Colour online)
 Three--dimensional density isosurface showing an
 off--centred vortex in a pancake--shaped
 condensate.}
\label{fig:vortex-init}
\end{figure}

The primary aim of this paper is to apply the ZNG model to quantised 
vortices at finite temperatures.
In a harmonically-trapped condensate at zero temperature a vortex will 
precess within the condensate, following a trajectory of constant radius, 
for which the energy remains constant. At finite temperature the 
presence of dissipation leads to the vortex minimising its energy by moving
towards the surface, where it eventually leaves the condensate and disappears.
This decay has been observed experimentally \cite{rosenbusch02,abo-shaeer02}, 
but the precise theoretical modelling of this process has proven
rather challenging.

Our approach should be contrasted with earlier work.
Dissipation based on the scattering of single-particle
excitations from the mean field potential at the vortex core was
considered by Fedichev and Shlyapnikov \cite{fedichev99} and was 
subsequently extended to treat vortex lattices 
\cite{zhuravlev01,fedichev02}. The problem of vortex nucleation
was then considered by Penckwitt {\it et al.}
\cite{Vortex_Penckwitt} using an approximation in which the
thermal cloud is treated as {\it static}.
Schmidt {\it et al.} \cite{schmidt03}
were the first to apply the classical field method to the
problem of vortex decay, while more recent applications of the 
method addressed vortex dynamics
in quasi-two-dimensional systems \cite{ENS_BKT,Classical_BKT}.
Duine {\it et al.} \cite{duine04} considered the 
dynamics of a straight vortex line using a closely-related
method based on a stochastic Gross-Pitaevskii equation 
\cite{Stochastic_Stoof} and derived a stochastic
equation of motion for the position of the vortex core; 
a similar equation was also used by Sasik {\it et al.}
to numerically simulate the motion of an isolated vortex in a uniform box 
with periodic boundary conditions  \cite{sasik00}.
Compared to the literature cited above,
our work represents the first microscopic simulations which
fully account for the dynamics of an inhomogeneous thermal cloud.

Our work, however, also has a second motivation.
Although the topic of quantised vorticity 
in superfluids is  interesting per se (as shown by the 
number of recent vortex experiments in atomic Bose--Einstein 
condensates),  it also has implications in the subject of
quantum turbulence \cite{Barenghi_Book}. Current
work on turbulent superfluid $^4$He and $^3$He-B, for
example, is 
concerned with the extent to which turbulence in these systems
differs from that found in ordinary classical
fluids \cite{tabeling98,vinen02,skrbek07,
barenghi08,Bradley,Tsubota-rotating,Tsubota-twisted,Lipniacki,
Kivotides-bundles}.
What makes quantum fluids attractive from
the point of view of understanding the principles of turbulence,
is the existence of various forms of dissipation which are
distinct from ordinary viscosity. At sufficiently
low temperatures, kinetic energy 
can be dissipated into sound waves, that is phonons \cite{vortex-heating}, 
via a Kelvin wave cascade \cite{Kelvin-cascade}
or via vortex reconnections \cite{pulses}. At higher temperatures,
the friction force \cite{bdv82}
between the superfluid and the normal fluid component
can change the nature of the turbulent kinetic energy cascade. For example,
in the classical turbulence scenario \cite{Frisch}, the 
Richardson--Kolmogorov inertial cascade is limited at large wavenumbers 
where viscous dissipation destroys the small scales, whereas in superfluid
$^3$He-B at relatively high temperatures
the friction can limit the inertial cascade at small
wavenumbers \cite{vinen-he3}. 

The mutual friction between a superfluid
and a normal fluid is one of the most intricate issues of superfluidity
\cite{sonin87}; in particular, the existence of a transverse force on
quantised vortices parametrised by a dimensionless temperature
dependence quantity called $\alpha'$ has been 
controversial \cite{iordanskii66,sonin75,sonin97,ao93}.
Studying vortex motion in an atomic BEC at finite temperatures, 
and  interpreting the results from the point of view of 
vortex dynamics, allows one to
compute the friction force directly from first principles.
This work on an atomic quantum fluid thus provides insights
into this important problem which cannot be obtained as readily
with superfluid helium.

This paper is structured as follows:
Sec.\ \ref{Theory} briefly reviews the ZNG theoretical model and numerical
implementation. Sec.\ \ref{Results} presents our main findings on vortex 
decay (Sec.\ \ref{Vortex_Decay}), highlighting the dependence of the 
friction coefficients on system parameters (Sec.\ \ref{Friction}), 
the role of collisions between the atoms (Sec.\ 
\ref{subsec:collisionless}), 
and the effect of thermal cloud rotation (Sec.\ \ref{Rotating_Cloud}).
Sec.\ \ref{Conclusions} presents some concluding remarks and briefly 
discusses the consequences of our analysis on the motion of vortex 
lattices.

\section{Theory}
\label{Theory}

\subsection{ZNG formalism}

Following Refs \cite{zaremba99} and \cite{jackson02a},
the second-quantised field operator 
$\hat{\psi} (\brt)$ can be split  into condensate and thermal 
contributions. Making use of Bose broken symmetry,
the full operator is written as
$\hat{\psi} (\brt)=\Psi(\brt)+\tilde{\psi} (\brt)$, where 
$\Psi(\brt)=\langle \hat{\psi} (\brt) \rangle$ is the condensate 
wavefunction (angular brackets denote an expectation value) and 
$\tilde{\psi}(\brt)$ is the noncondensate field operator. 
Starting with the Heisenberg equation of motion for $\hat{\psi}
(\brt)$, one eventually arrives at the following pair of
equations: 
\begin{eqnarray}
&& i\hbar \frac{\partial \Psi}{\partial t} = 
 \left( -\frac{\hbar^2 \nabla^2} {2m} + V +gn_c + 2g\tilde{n} - iR \right) 
 \Psi, \label{eq:GP-gen} \\
&& \hskip .15truein \frac{\partial f}{\partial t} + \frac{\bm p}{m} 
\cdot {\nabla} f 
 -{\nabla} U \cdot {\nabla_{\bm p}} f = C_{12} + C_{22}.
\label{eq:Boltz}
\end{eqnarray}
Eq.~(\ref{eq:GP-gen}) is a generalised Gross--Pitaevskii (GP)
equation for the condensate wavefunction $\Psi({\bm r}, t)$, and 
has been obtained in the so-called `Hartree-Fock-Popov'
approximation~\cite{Griffin_Popov}, whereby the static value of the 
`anomalous' average, which is responsible for certain
many-body
effects~\cite{Stoof_NIST,Proukakis_PRA,Shi_Griffin,Hutchinson_Review},
is ignored.
Eq.~(\ref{eq:Boltz}) is a 
Boltzmann equation for
the thermal cloud phase space density $f({\bm p},{\bm r},t)$,
with thermal energies
calculated in the Hartree-Fock approximation
\cite{Stringari_Review}. The 
condensate density is defined as $n_c = |\Psi|^2$, while the
thermal cloud density is obtained from $f$ by means of 
the momentum integral 
$\tilde{n}= \int d{\bm p}/h^3 \, f$. The mean field interactions
between atoms is parameterised by
$g=4\pi \hbar^2 a/m$, where $m$ is the atomic mass and $a$ is
the s-wave scattering length.
The thermal atoms experience an effective potential 
given by 
$U({\bm r})=V({\bm r})+2g(n_c+\tilde{n})$, where $V({\bm r})=m(\omega_{\perp}^2 r^2+\omega_z^2 z^2)/2$ 
represents the external trap. 

The terms involving $gn_c$ and $g \tilde{n}$ in Eq.\ (\ref{eq:GP-gen}) and ${\bf \nabla} U$ in Eq.\ (\ref{eq:Boltz}) represent mean field coupling
between atoms in the condensate and the thermal
cloud. This coupling is a source of dissipation for the system, and gives
rise, for example, to Landau damping of collective modes 
\cite{pitaevskii97,giorgini98,guilleumas00,jackson03}. 
The ZNG model also includes `collisional integrals' 
$C_{22}$ and $C_{12}$, which respectively denote
binary collisions between noncondensate atoms, and between condensate 
and noncondensate atoms. They are given by:

\begin{widetext}
\begin{equation}
 C_{22} = \frac{2 g^2}{(2 \pi)^5 h^7} \int d{\bm p_2} d{\bm p_3}
 d{\bm p_4} \, \delta(\bp+\bp_2-\bp_3-\bp_4) \delta(\epsilon+\epsilon_2
 -\epsilon_3-\epsilon_4) [(1+f)(1+f_2)f_3 f_4 - f f_2 (1+f_3)(1+f_4)],
\label{eq:c22}
\end{equation}
\begin{eqnarray}
 C_{12} &=&\frac{2 g^2 n_c}{(2\pi)^2 h^4} \int d {\bm p_2} d
{\bm p_3} d {\bm p_4} \, \delta(m\bv_c + \bp_2-\bp_3-\bp_4) 
 \delta(\epsilon_c+\epsilon_2-\epsilon_3-\epsilon_4) 
 [\delta(\bp-\bp_2)-\delta(\bp-\bp_3)-\delta(\bp-\bp_4)] \nonumber \\
 && \qquad \qquad \qquad \qquad \qquad 
 \times [(1+f_2)f_3 f_4 - f_2 (1+f_3)(1+f_4)],
\label{eq:c12}
\end{eqnarray}
\end{widetext}
where $f \equiv f (\bp, \br, t)$ and $f_i \equiv f (\bp_i, \br, t)$.
In the above expressions, delta functions enforce momentum and energy 
conservation in the 
collisions, where $\epsilon=p^2/(2m)+U$ is the thermal atom energy (in the Hartree-Fock limit),
$\epsilon_c=mv_c^2/2 + \mu_c$ is the local condensate energy, with 
$\bv_c=\hbar\,(\Psi^* \nabla \Psi - \Psi \nabla \Psi^*)/(2imn_c)$, and 
$\mu_c$ is the chemical potential. 

The $C_{12}$ term (\ref{eq:c12})
involves those collisions between condensate and thermal atoms which
lead to a transfer
of atoms between condensate and thermal cloud. This term is thus
related to the source term $-i R \Psi$ appearing in
Eq. \ (\ref{eq:GP-gen}) via 
\begin{equation}
 R (\brt) = \frac{\hbar}{2n_c} \int \frac{d\bp}{(2\pi\hbar)^3} \, C_{12}. 
\label{eq:rterm}
\end{equation}
If $R$ is positive (negative), there is
a net local flux of atoms {\it out of (into)} the condensate.

\subsection{Numerical methods}

The methods used for our numerical simulations are discussed more fully in 
Ref.\ \cite{jackson02a}, and are only briefly reviewed here. 

Firstly, we must generate a suitable initial state for the
simulations which consists of a condensate in equilibrium with
a thermal cloud at temperature $T$. The condensate wavefunction
can be obtained by an imaginary time propagation ($t \rightarrow-it$) 
of (\ref{eq:GP-gen}) with $R=0$. The thermal cloud density
$\tilde{n}=0$ is first set to zero, and an approximate thermal
cloud potential $U \simeq V+2gn_c$ is constructed from the
self-consistently determined condensate wavefunction. This
yields the initial thermal cloud density $\tilde{n}_0 (\br) = 
(1/\Lambda^3) g_{3/2} (z)$,
where $\Lambda=(2\pi\hbar^2/mk_B T)^{1/2}$ is the thermal de Broglie 
wavelength, $z(\br)=\exp\{\beta[\mu_c-U({\bm r})] \}$ is the local 
fugacity and $\mu_c$ is the condensate chemical potential. This
density is then used in (\ref{eq:GP-gen}) to obtain an improved
condensate wavefunction and the procedure is iterated until a
self-consistent solution of both the condensate and thermal
cloud densities is obtained. (For more details see \cite{jackson02a}.)

The vortex state of interest in our simulations can be obtained
from this equilibrium state by multiplying the condensate wavefunction 
by the phase factor $\exp[iS(\br)]$, where 
$S(\br)=\arctan((y-y_0)/(x-x_0))$ is the phase profile associated
with a straight vortex located at $(x_0,y_0)$. The GP equation is 
then solved again in imaginary time for a short period, until a vortex 
is fully formed in the condensate and most short time scale
transients in the initial configuration (for example, phonon
excitations) have damped out.

The state generated in this way is a quasi-equilibrium state
containing one vortex whose subsequent dynamical evolution
is of interest. The generalised Gross-Pitaevskii equation for the 
condensate wavefunction (\ref{eq:GP-gen}) can be solved readily
in real time using standard methods 
\cite{minguzzi04}, in our case a split-operator Fast Fourier Transform
approach on a 3D Cartesian grid. The thermal cloud 
is described by a swarm of classical test particles, which move in an 
external potential $U(\brt)$, and which, during a 
timestep, can collide with each other or with the condensate. 
The probabilities of particle collisions are chosen so that they
correspond to a Monte Carlo evaluation of the collision integrals in
(\ref{eq:c22}) and (\ref{eq:c12}). Together with the Newtonian
dynamics of the test particles, this procedure is equivalent
to solving the collisional Boltzmann equation (\ref{eq:Boltz}). The  
$C_{12}$ probabilities are then summed according to (\ref{eq:rterm}) to 
determine $R(\brt)$, which appears in the GP equation~(\ref{eq:GP-gen}).
Since $R(\brt)$  is a non-Hermitian term, it leads to change in the 
normalisation of the wavefunction, corresponding to condensate growth 
or loss, which is accompanied by a compensating removal or creation of 
thermal particles. An essential ingredient in the simulations is the
evaluation of the thermal cloud density $\tilde{n} (\brt)$, 
which appears in both (\ref{eq:GP-gen}) and (\ref{eq:Boltz}) 
(through the effective potential $U$). This is achieved by
appropriately binning the thermal particles and then convolving
the binned distribution with a 
Gaussian in order to obtain a smoothly varying potential.

\section{Results}
\label{Results}

The first simulation to be described is for a pancake-shaped 
condensate with $N=10^4$ $^{87} {\rm Rb}$ atoms with trap 
frequencies $\omega_{\perp}=2\pi \times 129\, {\rm Hz}$ and 
$\omega_z = \sqrt{8} \omega_\perp$. For these parameters the (ideal gas) 
critical temperature is $T_c = 177\, {\rm nK}$. This geometry has the 
advantage that the radius of the condensate in the axial direction is much
smaller than in the radial, so that the vortex remains relatively straight
throughout its motion. This simplifies the analysis considerably, as the 
vortex dynamics can be characterised by just its $x$ and $y$ coordinates. 

\subsection{Vortex decay}
\label{Vortex_Decay}

Following the initial preparation phase, the vortex is  centred on 
$x_v(0) \simeq 1.3 a_{\perp}$, $y_v(0)=0$, as illustrated 
in Fig.~(\ref{fig:vortex-init}), where $a_{\perp}=\sqrt{\hbar/m 
\omega_{\perp}}$ is the harmonic oscillator length in the radial 
direction. The subsequent vortex position 
${\bf r}_v (t) = (x_v(t),y_v(t))$ is  tracked by finding the
corresponding local density minimum in the $z=0$ plane by
means of a quadratic interpolation  
between grid points. This procedure breaks down when the vortex leaves 
the bulk of the condensate into the very low densities beyond the edge, so 
no results are shown when this happens. This condition can
be taken as the point
at which the vortex ``disappears'' from the condensate, and would 
 correspond in the experimental context to the density
contrast of the vortex being below the detection limit. 
Since the initialisation process induces
a centre-of-mass motion of the condensate,
the following analysis depicts the vortex position
relative to the centre-of-mass position.

Simulations of the GP equation for $T=0$ reveal that the vortex 
precesses in a circular path around the condensate, following a trajectory
of constant energy as would be expected for a non-dissipative system. This 
well-known precessional behaviour can be understood as arising from the 
non-uniform density of the condensate, which means that the energy of the 
vortex is a function of its radial position. Hence there is an 
effective Magnus force, proportional to the gradient of the energy 
and directed radially, with in turn induces the 
azimuthal vortex motion \cite{jackson99}. This can be described
quantitatively by means of a time-dependent variational method 
\cite{svidzinsky00,lundh00,fetter01}.
In addition to the motion described above, the acceleration 
experienced by the vortex in its
circular trajectory can in principle lead to the emission of
sound waves. However, for the harmonic confinement being
consider, any emission of sound waves is followed by
reabsorptioon, and no net dissipation occurs. There is
nevertheless some modulation of the vortex trajectory arising
from the dynamical interaction between of the vortex and sound
waves \cite{Parker_PRL_2004}.

\begin{figure}[t]
\centering \scalebox{0.5}
 {\includegraphics{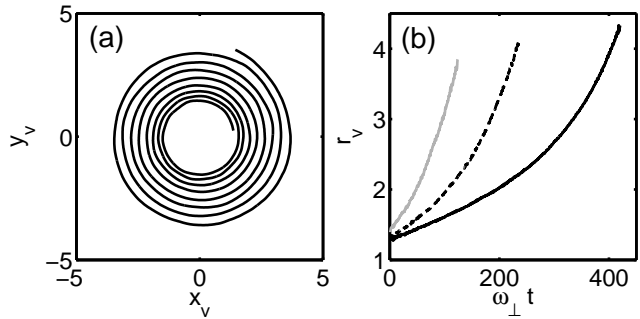}}
 \caption{(a) Position of the vortex $(x_v, y_v)$ (in units
 of $a_\perp = \sqrt{\hbar/m\omega_\perp}$), as it 
 spirals out of the condensate at $T=0.7 \, T_c$. 
 (b) Radial position of the  
 vortex $r_v=\sqrt{x_v^2+y_v^2}$ as function of time (in units of  
 $\omega_{\perp}^{-1}$), for $T=0.5\, T_c$ (solid black line), $T=0.6\, T_c$ (dashed
 black), and $T=0.7\, T_c$ (solid grey).}
\label{fig:vortpos-xyrt}
\end{figure}

In contrast to this dissipationless motion, the vortex can lose
energy at finite temperatures due to 
interactions with the thermal component, and as a result, it moves 
radially towards the condensate
edge. The resulting spiral trajectory is illustrated in  Fig.\ 
\ref{fig:vortpos-xyrt}(a) for a temperature of $T=0.7\, T_c$. Fig.\ 
\ref{fig:vortpos-xyrt}(b) shows the outward relaxation of the radial 
position $r_v = |{\bf r}_v| = \sqrt{x_v^2 + y_v^2}$ as 
a function of time. The three curves represent different temperatures, and
demonstrate, as one would expect, that the relaxation rate 
increases with temperature.

\begin{figure}[b]
\centering \scalebox{0.25}
 {\includegraphics{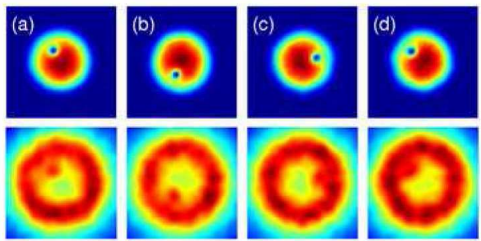}}
 \caption{(Colour online) Density cross-sections of the condensate (top row) 
 and thermal cloud
 (bottom row) at $z=0$ for $T=0.7\, T_c$, and at the times (a) 
 $\omega_{\perp} t=6$, (b) 12 (c) 18, and (d) 24. The colours
 range from brown/red (high density) to dark blue (low density), with 
 different scales for the condensate and thermal cloud densities.}
\label{fig:vortcomp}
\end{figure}

As well as monitoring the vortex position, it is instructive to study 
the evolution of the densities during the simulation. Fig.\ 
\ref{fig:vortcomp} shows the condensate (top images) and thermal cloud 
densities (bottom) (cross-sections at $z=0$) for various 
times, and for a temperature of $T=0.7\, T_c$. The density
dip at the vortex core (dark blue) is evident in the condensate 
images (top), and executes the spiral motion of 
Fig.\ref{fig:vortpos-xyrt}(a). 
The bottom images show the thermal cloud density; as this is much 
smaller than the condensate density, a different colour scale is used 
here for clarity. The thermal cloud density is
largest near minima of the effective potential $U(\brt)$,
leading in the present situation to the
circular ring of peak thermal cloud density (red) at the edge of 
the condensate where $n_c \rightarrow 0$. However, the thermal
cloud also tends to ``fill in'' the vortex core since the lower
condensate density at this position gives rise to a dip in
the effective potential. The resulting peak in the thermal cloud
density tends to follow the core as it precesses and spirals
out. We note that the 
 stabilisation of the vortex by the thermal cloud suggested
in \cite{Virtanen} is not observed. It arises when the thermal
cloud is treated as {\it static} \cite{Virtanen}, but does not
occur when the dynamics of the thermal cloud is taken into
account.

\begin{figure}[ht]
\centering \scalebox{0.43}
 {\includegraphics{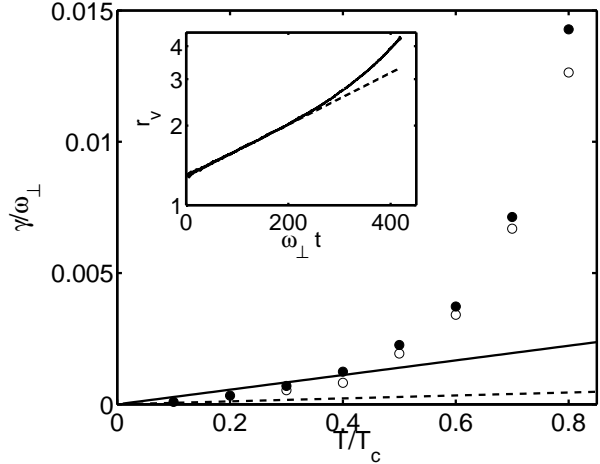}}
 \caption{(Inset) Lin-log plot of the vortex radial position as a function of 
 time, for $T=0.5\, T_c$ (solid line). The 
 dashed line is an exponential fit, $r_v(t)=r_0 e^{\gamma t}$, to the data 
 over $0\leq t \leq 50$. The main figure plots the resulting values of  
 $\gamma$ for for an initial vortex position of $r_0 \simeq 1.3$ 
 (solid circles) and $r_0 \simeq 0.65$ (open circles). For comparison, the 
 solid and dashed lines plot the results of FS \cite{fedichev99} and DLS 
 \cite{duine04} respectively.}
\label{fig:gamma1}
\end{figure}

The radial position of a vortex close to the trap centre
exhibits a near-exponential growth in time. This is evident from 
the inset of Fig.\ \ref{fig:gamma1}, which plots the time evolution 
of $r_v$ on a logarithmic-linear scale for $T=0.5\, T_c$. 
Deviations from this simple exponential behaviour are observed
at later times when the vortex approaches the edge of the
condensate; similar behaviour is found for other temperatures.
To quantify the relaxation we thus fit $r_v(t)$
over $0 \leq t \leq 50$ to the function $r_v (t) = r_0 \exp (\gamma t)$. 
The resulting values of $\gamma$ are plotted in the main panel of 
Fig.~\ref{fig:gamma1} with the black circles.

The departure from a pure exponential behaviour can be interpreted as a
position-dependent relaxation rate $\gamma$ which is enhanced by
the local maximum in the thermal atom density near the edge of the 
condensate (see Fig.\ \ref{fig:vortcomp}). To check this
interpretation, we performed simulations in which the vortex
starts nearer to the centre, e.g.\ at $r_0 \simeq 0.65
a_{\perp}$, instead of the initial position $r_0 \simeq
1.3 a_{\perp}$ used earlier.
The resulting values of $\gamma$ are plotted 
as the open circles in Fig.~\ref{fig:gamma1}, showing smaller values
consistent with the lower thermal cloud density closer to the
centre of the trap. However, 
the difference is small, showing that the exponential decay approximation 
is good until quite near to the condensate edge, where the larger thermal 
cloud density leads to a more rapid relaxation 
(inset of Fig.~\ref{fig:gamma1}).

It is of interest to compare our results to 
existing analytical predictions  obtained by Fedichev and 
Shlyapnikov (FS) \cite{fedichev99}, and Duine, Leurs and Stoof (DLS)
\cite{duine04}. Both FS and DLS actually quote the 
timescale $\tau$ for the decay of a vortex from position
$r_{\rm min}$ to $r_{\rm max}$, but this can be converted to a decay
rate using $\tau =  \gamma^{-1} \ln (r_{\rm max}/r_{\rm min})$. 
In both of these works the decay of the vortex is found to be 
exponential, with a rate $\gamma \propto T$ 
but with different proportionality coefficients due to the 
different  approximations made in the two theories (see below 
for the role of the noise term in the DLS analysis). 
These rates are displayed in Fig.\ \ref{fig:gamma1} by solid (FS) 
and dashed lines (DLS). It is apparent that the rates found by
FS are comparable to ours at the lower temperatures, but differ
significantly at higher $T$ due to the stronger (approximately
quadratic) temperature dependence found in our simulations.

FS assumed a  uniform condensate in a cylindrical container, and modelled
the decay solely as the result of mean-field interactions.
The DLS study on the other hand, includes the important $C_{12}$ 
collisional coupling between the condensate and thermal cloud.
With the aim of obtaining analytical results, they approximate the 
condensate density profile by a Gaussian which is reasonably accurate 
for the most relevant region near the center of the trap 
\cite{Stoof_Private}. In this regard, it should be noted that such an
approximation has been shown to produce correct results for the
frequencies of collective modes even for Thomas-Fermi condensates 
\cite{Duine_Stoof}.
%
Our present simulations, which include both mean field and
collisional coupling mechanisms, enable us to assess their relative 
importance, which will be discussed in Sec.\ \ref{subsec:collisionless}.
More importantly, our simulations differ from these approaches in that 
they are actually performed for a dynamical thermal cloud, with
both the condensate and thermal cloud densities determined 
self-consistently during the simulations. 

For completeness, we should however make two additional remarks
regarding the DLS approach. Firstly, in their preceding work 
\cite{Duine_Stoof}, Duine and Stoof argued 
that enhanced damping of collective modes at higher temperatures could be
related to the position dependence of the self-energy (and hence of the
damping term $-iR\Psi$), which was ignored in their analytical treatment
based on a volume average over the size of the condensate. In the present
context, this could partly account \cite{Duine_Private} 
for a deviation of the computed damping rate $\gamma/\omega_{\perp}$ 
at higher temperatures from the linear behaviour seen in 
Fig.~\ref{fig:gamma1}. Although such simulations have not been 
performed to date, the resulting corrections are likely to be smaller 
than the observed disagreement, whose origin we believe lies primarily 
in the dynamics of the thermal cloud.

On the other hand, the DLS analysis is actually more 
general than ours in that it contains an additional noise term in the 
equation of motion of the vortex. This provides stochastic
``kicks" to the vortex and the ensuing Brownian motion allows
for the migration of a vortex away from the centre of the trap.
Were it not for the noise, a centred vortex would have an infinite 
lifetime. This would be the case in an exact application of the
ZNG theory. However, its numerical implementation in terms of
discrete test particles does introduce statistical fluctuations
in the thermal cloud density which plays the role of noise. Thus
we indeed find in simulations of a centred vortex a finite,
albeit long, lifetime ($\omega_{\perp} \tau \simeq 1000$
for the relatively high temperature of $T=0.7\, T_c$ as compared
to $\omega_{\perp} \tau \simeq 120$ for $r_v(0) \simeq 1.3
a_{\perp}$). This long lifetime, however, should not be taken
seriously since it depends on the actual number of test
particles used in the simulations. The simulation nevertheless makes 
clear that this `numerical noise' is of secondary importance at larger 
radii where the direct coupling to the thermal cloud is the dominant
dissipative effect. Although it is something to be checked, it
is unlikely that the stochastic term makes a significant 
contribution to the spiralling out of a vortex when it is located 
far from the trap centre.

Finally, it is worth remarking that our computed decay rates 
(from approximately $0.5~\rm s^{-1}$ to $3~\rm s^{-1}$ in the range
$T/T_c=0.4$ to $0.6$) 
are in order-of-magnitude agreement with the decay rates
observed for a vortex lattice \cite{abo-shaeer02}
(approximately $0.3~\rm s^{-1}$ to $3~\rm s^{-1}$ over the same relative 
range). 

\subsection{Friction coefficients} 
\label{Friction}

In order to further understand the origin of this exponential decay, it 
is instructive to consider the two--fluid hydrodynamics
model used to describe superfluid liquid helium \cite{donnelly-book}.
In this context, dissipation arises from the interaction between the
quantised vortices and the
thermal excitations (phonons and rotons) which form the normal
fluid \cite{hall56,bdv82}.  Since the radius of the superfluid vortex core 
is much smaller than the typical separation between vortices or any other
length scale of interest in the flow, the vortex is described
in parametric form as a three-dimensional space curve 
${\bf s} \equiv {\bf s} (\xi, t)$, where $\xi$ is the arclength.
The resulting equation of motion \cite{schwarz88} is:
\begin{equation}
 \frac{d {\bf s}}{d t} = {\bf v}_s + {\bf v}_i + \alpha {\bf s}' \times
 ({\bf v}_n-{\bf v}_s - {\bf v}_i) 
 - \alpha' {\bf s}' \times 
 [{\bf s}' \times ({\bf v}_n - {\bf v}_s - 
 {\bf v}_i)],
\label{eq:schwarz}
\end{equation}
where ${\bf s'}=d{\bf s}/{d \xi}$ is the unit tangent along the vortex
at the position ${\bf s}$, ${\bf v}_s$ is any
imposed superfluid velocity, ${\bf v}_n$ is the normal fluid velocity, and
${\bf v}_i$ is the self--induced velocity of
the vortex arising from its own curvature, the presence of other 
vortices, and  any inhomogeneity of the fluid. The first two terms 
on the right hand side of Eq.~(\ref{eq:schwarz})
state that at $T=0$ the vortex is advected by the local superflow 
${\bf v}_s+{\bf v}_i$. The remaining
two terms reflect the fact that, at nonzero $T$,
the normal fluid streaming past the vortex
core exerts a force per unit length whose intensity is controlled by the
temperature--dependent friction coefficients $\alpha$ and $\alpha'$
\cite{bdv82,db98,Bevan}. 

In turbulent helium, the friction coefficients $\alpha$ and 
$\alpha'$ play a key role because they account for the mutual 
friction between the superfluid and the normal fluid and control 
the transfer of energy between the two fluids at various length scales 
and hence the nature of the inertial range cascade \cite{vinen02,Hulton}. 
The first coefficient, $\alpha$, describes dissipative effects
and leads, for example, to the shrinking \cite{bdv82} of a vortex ring  
or the damping of a Kelvin wave \cite{bdv85} in a normal fluid at rest.
In the case of rotating helium, $\alpha$ determines the 
attenuation of second sound waves, so it allows the experimentalist
to determine the density of vortex lines \cite{bgs06}.
The second coefficient,
$\alpha'$, is not dissipative, and, in rotating helium, splits
a second sound resonance (besides the classical rotational splitting)
in a suitably designed cavity \cite{Lucas}.
It has been argued \cite{finne03,finne05} that the ratio of inertial 
and dissipative forces,
which is called the Reynolds number in the case of ordinary turbulence
\cite{Frisch}, is simply $\alpha/(1-\alpha')$ in the case of quantum 
turbulence.

In our case of 
a pancake-shaped condensate the vortex line remains approximately straight, so
we can then replace ${\bf s}$ by the vortex position ${\bf r}_v$, while
${\bf s}'=\hat{\bf z}$. Both the condensate and thermal cloud are stationary,
and so ${\bf v}_s={\bf v}_n=0$. This just leaves ${\bf v}_i$, which in this 
case corresponds to the azimuthal vortex motion induced by the inhomogeneity 
of the condensate at zero $T$ (hence, ${\bf v}_i = v_i \hat{\bm{\phi}}$).
Rewriting Eq.\ (\ref{eq:schwarz}) in cylindrical polar coordinates gives
for the azimuthal component
\begin{equation}
 \omega_v = (1-\alpha') \frac{v_i}{r_v},
\label{eq:alpha-pr}
\end{equation}
where $\omega_v = \dot{\phi_v}$ is the vortex precession frequency at 
finite $T$ and the dot denotes a time derivative. 

To a first approximation we ignore the 
mutual friction coefficient $\alpha'$. We then find 
$\omega_v = v_i/r_v$, and thus obtain for the radial component
\begin{equation}
 \frac{d r_v}{d t} = \alpha \omega_v r_v \;.
 \label{eq:radial-decay}
\end{equation}
If $\alpha$ and $\omega_v$ are constant, then one simply recovers the
exponential behaviour discussed in the previous section,
with $\gamma = \alpha \omega_v$. 

In actual fact, $\alpha$ and $\omega_v$ would not be expected to be 
constant during the course of the vortex decay. The condensate and 
thermal cloud are both non-uniform (implying a dependence of $\alpha$ 
on position), while $\omega_v$ is only constant near  the 
centre, and increases as the vortex approaches the edge \cite{jackson99}. 
However, from the evidence of Fig.\ \ref{fig:gamma1} (inset), 
Eq.~(\ref{eq:radial-decay}) is a good approximation for times when 
the vortex is close to the centre where the densities are relatively 
uniform. 

\begin{figure}[t]
\centering \scalebox{0.45}
 {\includegraphics{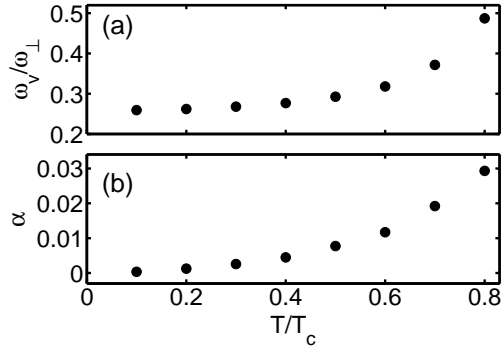}}

 \caption{(a) Precession angular frequency of the vortex, $\omega_v$,
 as a multiple of the radial trap frequency $\omega_{\perp}$. 
 (b) Friction coefficient $\alpha$ as a function of relative
 temperature $T/T_c$, 
 for $N=10^4$ and a pancake trap geometry 
 $\omega_z/\omega_{\perp}=\sqrt{8}$.
}
\label{fig:alpha1}
\end{figure}

Values of $\alpha$ can therefore be estimated from $\gamma$ in Fig.\ 
\ref{fig:gamma1}. The other necessary ingredient is $\omega_v$, 
which can be calculated as a function of time using $\omega_v = x_v \dot{y}_v - 
y_v \dot{x}_v$, where numerically the time derivatives are calculated 
using central differences. Due to errors in locating the vortex position, 
there are large fluctuations in the calculated $\omega_v$.
To obtain the corresponding values we thus average results over $2-3$ 
orbits of the vortex, corresponding to times $0<\omega_{\perp}t<50$.
The resulting mean values are plotted in Fig.\ \ref{fig:alpha1} (a). 
These results are also used for calculating the values of $\alpha$ 
presented in Fig.\ \ref{fig:alpha1} (b). The observed increase in 
$\alpha$ with rising $T$ is similar to the behaviour
observed in liquid $^3$He \cite{Bevan} and $^4$He \cite{bdv82,db98},
and is consistent with results obtained using the related but simpler
phenomenologically damped GP equation\cite{eniko}.

Let us now investigate the role of the $\alpha'$ coefficient in
Eq.\ (\ref{eq:schwarz}).
Experiments in liquid helium show a small effect, and 
there has been some controversy in the literature about this transverse
component of the friction force \cite{iordanskii66,sonin75,sonin97,ao93}.
This issue can {\it be addressed} within our simulations by comparing the 
``dynamic'' thermal cloud frequency $\omega_v$ to a ``static'' value 
$\omega_{\rm st}$ found using a GP simulation in which the
dynamics of the thermal cloud is ignored, that is, it retains
its initial equilibrium form for all times. In  this {\it static
thermal cloud approximation}, the thermal cloud exerts a
time-independent mean-field potential on the condensate, and
its effect can be identified with the $v_i/r_v$ term in 
Eq.~(\ref{eq:alpha-pr}), giving
the simple relation for $\alpha' = (1-\omega_v/\omega_{\rm st})$. We find
that $\omega_v$ and $\omega_{\rm st}$ are equal to within $2-3\ \%$,
showing that the changes observed in $\omega_v$ in Fig.\ \ref{fig:alpha1}
are almost entirely due to the effects of the static thermal cloud potential
on the condensate density profile, and not to ``real'' dynamical effects of
finite temperatures.  
The errors in measuring the vortex precession frequency are such that
we cannot confidently extract a value for $\alpha'$, although our simulations indicate
that $|\alpha'| < 0.02$ throughout the measured temperature range
$0 < T/T_c < 0.8$. This
agrees with recent results of Berloff and Youd~\cite{berloff07} obtained 
by means of classical field theory.

\begin{figure}[ht]
\centering \scalebox{0.55}
 {\includegraphics{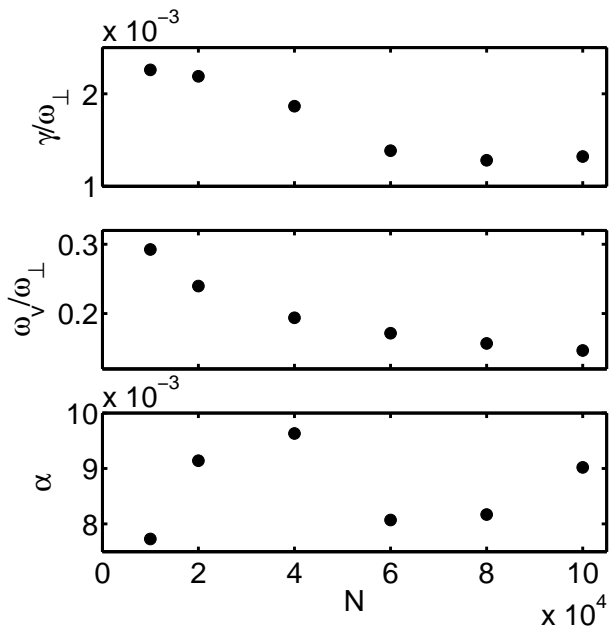}}
 \caption{(a) Decay rate $\gamma$, (b) precession frequency $\omega_v$,
 and (c) $\alpha$, as a function of total number of atoms $N$, for 
 temperature $T/T_c=0.5$ and a pancake geometry.}
\label{fig:natom1}
\end{figure}

We also explore the dependence of these parameters on the total number of atoms, $N$. 
Fig.\ \ref{fig:natom1} shows results for $N$ between $10^4$ and $10^5$, and
a fixed value of $T=0.5 \, T_c$. The decay rate $\gamma$, plotted in 
Fig.\ \ref{fig:natom1} (a), tends to decrease with increasing $N$. 
The vortex precession frequency shown in Fig.~\ref{fig:natom1} (b)
also has this decreasing trend, in agreement with what one 
would expect from GP solutions. These decreasing trends approximately 
cancel  when calculating $\alpha=\gamma/\omega_v$, leading to 
variations having no clear dependence on $N$. The scatter
reflects the uncertainty in the calculation of $\alpha$ and suggests 
that this parameter is approximately independent of atom number.

\subsection{Collisionless simulations}
\label{subsec:collisionless}

\begin{figure}[t]
\centering \scalebox{0.43}
 {\includegraphics{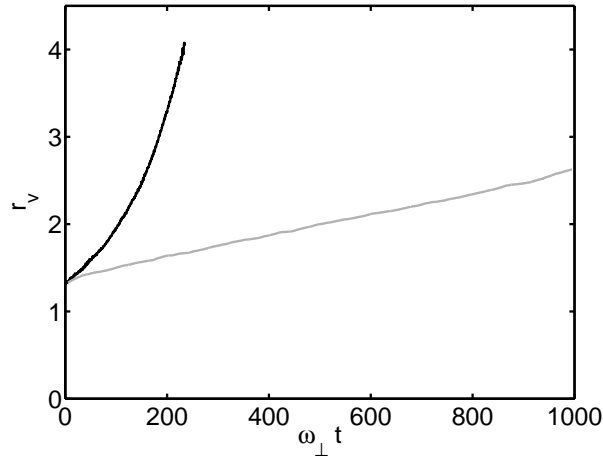}}
 \caption{Radial position of the vortex {\it vs.\ }time for $T = 0.6\, T_c$, 
 and collisional (black line), and collisionless (grey line) simulations.}
\label{fig:coll-less}
\end{figure}

In general, damping in the ZNG formalism arises from the coupling 
of the condensate to the thermal cloud by means of mean field 
interactions and $C_{12}$ collisions. In order to explore the relative 
importance of these two contributions, we have performed simulations 
where collisions are not present, so $C_{12}=C_{22}=0$ in 
(\ref{eq:Boltz}), and the only source of dissipation is mean-field
coupling. Physically, this dissipation is a form of Landau
damping whereby the motion of the vortex core through the
thermal cloud generates thermal excitations~\cite{jackson03}. In 
Fig.\ \ref{fig:coll-less} the radial position for the collisionless 
simulation at $T=0.6\, T_c$ is shown as the grey line, and compared to the 
collisional result in black. Without collisions one obtains a much slower
decay, highlighting the crucial role of collisional damping in the 
simulations, a conclusion which has been numerically verified over a 
broad range of temperatures. The near linear variation of
$r_v$ with time seen in Fig.\ \ref{fig:coll-less} indicates that
an exponential function would be a poor fit to the data in this
case. In view of Eq. (\ref{eq:radial-decay}), it would appear
that the mean-field contribution to $\alpha$ must be a
decreasing function of $r_v$, but we have no simple explanation
for this.

\subsection{Rotating thermal clouds}
\label{Rotating_Cloud}

We also consider the case where, instead of being stationary initially, 
the thermal cloud undergoes solid-body rotation around the $z$-axis with 
angular frequency $\Omega_{\rm th}$. The thermal cloud velocity at the 
vortex core is then  ${\bf v}_n = 
\Omega_{\rm th} r_v \hat{\bm{\phi}}$. Hence $\Omega_{\rm th}>0$ 
represents rotation in the same sense as the vortex precession, while
$\Omega_{\rm th}<0$ indicates rotation in the opposite sense. Using Eq.\ 
(\ref{eq:schwarz}) 
then yields:
\begin{equation}
 \omega_v = (1-\alpha') \frac{v_i}{r_v} + \alpha' \Omega_{\rm th},
\end{equation}
and,
\begin{equation} 
 r_v = r_0 \, e^{\alpha (\omega_v - \Omega_{\rm th}) t}.
\label{eq:vortpos-rot}
\end{equation}

\begin{figure}[t]
\centering \scalebox{0.35}
 {\includegraphics[angle=270]{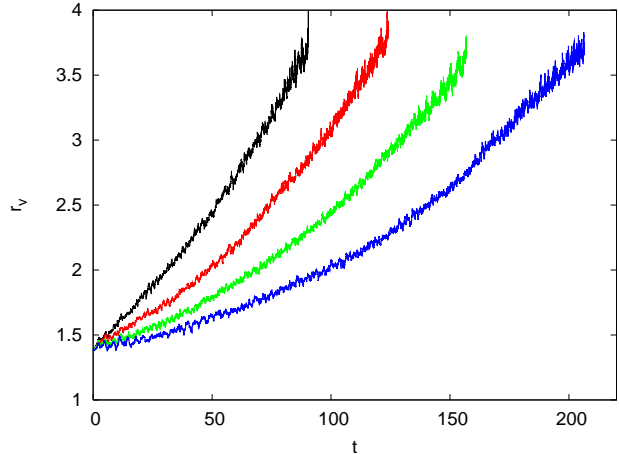}}
 \caption{(Colour online).
 Time evolution of vortex radial position for $T=0.7\, T_c$ and
 a rotating thermal cloud. The different curves represent varying thermal
 cloud rotation rates, with $\Omega_{\rm th} = -0.2$ (black, top), 
 $\Omega_{\rm th}=0$ (red), $\Omega_{\rm th} = 0.2$ (green), and 
 $\Omega_{\rm th}=0.37$ (blue, bottom).}
\label{fig:rot-therm}
\end{figure}

For these simulations we start with the equilibrium condensate and thermal 
cloud distributions evaluated at $T=0.7\, T_c$ and $\Omega_{\rm th}=0$. 
A rigid body rotation of the thermal cloud is then imposed by adding
${\bf v}_n = \Omega_{\rm th} r_v \hat{\bm{\phi}}$ to each atom's velocity.
It should be noted that the
thermal cloud will now no longer be in ``equilibrium'' since there is a 
centrifugal effect which tends to expand the cloud. This
initial outward expansion leads to an oscillation
in the radial direction and, since angular momentum is
conserved, a corresponding oscillation of the cloud's angular
velocity.

\begin{figure}[b]
\centering \scalebox{0.43}
 {\includegraphics{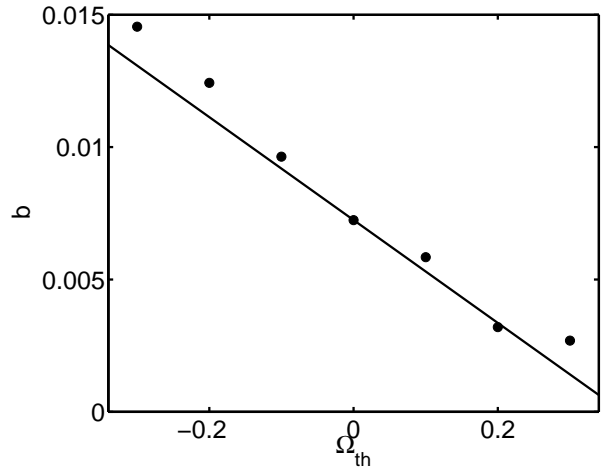}}
 \caption{Value of decay rate $\gamma$ for different rotation rates of the 
 thermal cloud, $\Omega_{\rm th}$. The solid line plots the function 
 $\gamma=\gamma_0 (1-\Omega_{\rm th}/\omega_{v0})$ (where $\gamma_0$ and
 $\omega_{v0}$ are the decay rate and the precession angular velocity
 respectively for a non-rotating thermal cloud), which is the expected
 dependence from Eq.\ (\ref{eq:vortpos-rot}).
}
\label{fig:rotation}
\end{figure}

The vortex decay curves are plotted in Fig.~(\ref{fig:rot-therm}) for different
$\Omega_{\rm th}$, and show that the decay rate increases for rotations 
opposite to the vortex precession direction ($\Omega_{\rm th}=-0.2$), but
decreases when they rotate in the same direction ($\Omega_{\rm th}=0.2$ and
$\Omega_{\rm th}=0.37$). This is consistent with the expected behaviour 
from Eq.\ (\ref{eq:vortpos-rot}). To study the problem more quantitatively,
we again fit exponentials of the form $a e^{bt}$ to the decay curves over 
$0 \leq \omega_{\perp} t \leq 50$. The values of $b$ for the different 
$\Omega_{\rm th}$ are plotted in Fig.\ \ref{fig:rotation}. The straight 
line is the expected result from (\ref{eq:vortpos-rot}), 
$b=\alpha (\omega_v - \Omega_{\rm th})$,
where $\alpha$ and $\omega_v$ are taken from the results of the 
$\Omega_{\rm th}=0$ simulations found earlier. Our results are
in quite good agreement with this behaviour, although some small 
discrepancies are apparent.
These could be due to the oscillations in $\Omega_{\rm th}$ noted earlier,
whereas Eq.~(\ref{eq:vortpos-rot}) assumes that $\Omega_{\rm th}$
is strictly time-independent throughout the precessional motion
of the vortex.

\section{Conclusions}
\label{Conclusions}

In summary, we have studied the finite temperature dynamics of a 
single vortex in a partially-condensed ultra-cold Bose gas. Our 
methodology and detailed simulations provide several advantages over 
previous studies.
Firstly,  our simulations include the full effects of the
trapping potential through the self-consistent determination of
condensate and thermal cloud in the initial state. This results
in a more realistic model as compared to those using the
approximation of uniform
densities~\cite{fedichev99,Vortex_Penckwitt,duine04}. The
inclusion of non-uniform densities accounts more realistically
for both the dynamics of the vortex and the positional dependence 
of the dissipation.
Secondly, our model includes both mean-field \cite{fedichev99} and 
collisional \cite{duine04}
damping, and allows us to compare the relative importance of the two 
mechanisms. Thirdly, the thermal cloud is not assumed to be
static as in earlier treatments, but is treated dynamically on
the same footing as the condensate. This more refined treatment
negates suggestions that the thermal cloud can act as a ``pinning
potential'', stabilising the vortex~\cite{Virtanen}. In addition, 
it has allowed us to observe the change in damping when 
the thermal cloud is moving relative to the condensate as, for
example, when undergoing a rotation. 

We also compared our results for the vortex relaxation rate to those of 
other studies and found some significant differences,
particularly with regard to its temperature dependence.
Furthermore, by comparing the trajectory of vortices with the 
predictions of phenomenological vortex dynamics equations, we
were able determined the mutual friction coefficients from first
principles.

\begin{figure}[h]
\centering \scalebox{0.2}
 {\includegraphics{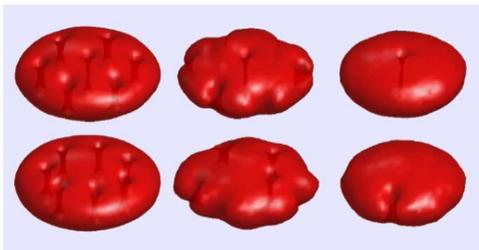}}
 \caption{(Colour online). Vortex lattice decay at $T=0.7~T_c$.
 The top row shows the time evolution (left to right)
 of a lattice with a vortex initially
 at the centre; the bottom row shows the evolution of a lattice with
 the same number of vortices but having a ring configuration.}
 \label{fig:lattice}
\end{figure}

Our approach can also be extended to study the role of a dynamical 
thermal cloud on vortex lattice dynamics 
\cite{VortexLattice_1,VortexLattice_2,VortexLattice_3},
thereby complementing and extending existing work 
\cite{zhuravlev01,fedichev02,Vortex_Lobo,Vortex_Penckwitt,Vortex_Tsubota,
Vortex_Simula,Virtanen,Vortex_SGPE}.
To illustrate this possibility, we conclude by briefly
reporting on some preliminary results for the decay of vortex
lattices. We consider the evolution of
two different vortex lattice configurations shown in 
Fig.~\ref{fig:lattice}. Both vortex arrays initially contain
seven vortices (left images), however they differ in the way the 
vortices are arranged. The first array (top images) consists of one
vortex at the centre of the condensate and a ring of six vortices 
around it, whereas the second array (bottom) consists of a ring of 
seven vortices with no central vortex. These arrays rotate in
the laboratory frame, and at finite temperatures, the effects of
dissipation lead to the gradual disappearance of the off-centred 
vortices, one by one. For simulations performed at $T = 0.7T_c$, 
we find that the array with 
all vortices arranged in a ring decays faster: after the first six 
vortices have decayed, the system is left with a single off-centre  
vortex which moves relatively rapidly to the edge and disappears. This 
whole evolution occurs on a time scale $\omega_{\perp}t\approx
150$. The decay rate is in order-of-magnitude agreement with 
measurements performed with a bigger lattice \cite{abo-shaeer02}
after the latter are extrapolated to our value of $T/T_c$.

In contrast to this, the lattice with the central vortex
reaches a point where a single metastable central vortex
remains after the other six have been shed.
This vortex also eventually decays, but our simulations suggest 
that the decay occurs on a much longer timescale. Some numerical
experiments we have performed for configurations with no initial 
central vortex have exhibited a similar metastable behaviour. If, 
during the initial part of the evolution (in which the vortices move
irregularly), a vortex ends up sufficiently close to the centre,
it can become ``stuck" near the centre while the other vortices
are shed. These observations are in agreement
with reports \cite{bretin03} that the decay time of the
last vortex is much longer than that of the initial vortex array.

We stress, however, that a more accurate treatment of the
evolution of a metastable central 
vortex requires the explicit inclusion of stochastic noise to
provide a``kick'', as discussed in~\cite{duine04,sasik00}.
Such a term was included in a recent 
discussion~\cite{Vortex_SGPE}, however the thermal cloud was
still treated as static. A combination of the stochastic 
Gross-Pitaevskii equation and the quantum Boltzmann equation may 
thus be needed to provide a more complete description of this
particular situation. However, we expect that most other cases
can be modelled extremely well by the Zaremba - Nikuni- Griffin 
approach. In particular, it would of interest to see if more
detailed calculations of vortex lattice decay would be consistent
with experimental observations~\cite{abo-shaeer02}. 
Other applications might include the study of vortex lattice
excitations (Tkachenko modes) \cite{Tkachenko}, and the dynamics 
of bent vortices in elongated condensates~\cite{rosenbusch02}.

\bigskip

This work was funded by EPSRC grant EP/D040892/1
(BJ, NPP, CFB) and by NSERC of
Canada (EZ).

\end{document}